\documentstyle[numreferences]{kluwer}     
\begin{opening}
  \title{On The Even CAR~Algebra} 
  \author{Carsten \surname{Binnenhei}\thanks{%
      Supported by the Deutsche Forschungsgemeinschaft (Sfb 288
      ``Differentialgeometrie und Quantenphysik'')}}
  \institute{Institut f\"ur Theoretische Physik\\Freie Universit\"at
    Berlin, Germany\\E--mail: binnenhe@physik.fu-berlin.de}  
  \date{April 1996}
\end{opening}
\input amssym.def
\newtheorem{prop}{PROPOSITION}

\begin{document}
\begin{abstract}
  We exhibit a family of *-isomorphisms mapping the CAR algebra onto
  its even subalgebra.
\end{abstract}

In 1970 E.~St\o rmer proved that the even subalgebra of the CAR
algebra over an infinite dimensional separable Hilbert space is UHF of
type $2^\infty$, hence *-isomorphic to the CAR algebra itself
\cite{S}. But it seems to be unknown that such isomorphisms have a
nice and simple realization in terms of Bogoliubov endomorphisms with
``statistical dimension'' $\sqrt{2}$.

Bogoliubov endomorphisms are conveniently described using Araki's
``selfdual'' CAR algebra formalism \cite{A}. Let ${\cal K}$ be an
infinite dimensional separable complex Hilbert space, equipped with a
complex conjugation $k\mapsto k^*$, and let ${\cal C(K)}$ denote the
unique (simple) C*-algebra generated by ${\bf1}$ and the elements of
${\cal K}$, subject to the anticommutation relation
$$k^*k'+k'k^*=\langle k,k'\rangle {\bf1},\quad k,k'\in{\cal K}.$$
${\cal K}$ will henceforth be viewed as a subspace of ${\cal C(K)}$
(but we caution the reader that the Hilbert space norm and the C*-norm
differ on ${\cal K}$). Each isometry $V$ on ${\cal K}$ that commutes
with complex conjugation extends to a unique unital *-endomorphism
$\varrho_V$ of ${\cal C(K)}$. Such $V$ has a well-defined Fredholm
index $\mbox{ind\,}V=-\dim\ker V^*$, and the statistical
dimension of $\varrho_V$ (i.e.~the square root of the index of the
inclusion $\varrho_V({\cal C(K)})\subset{\cal C(K)}$) was shown to
equal $2^{-\frac{1}{2}\mbox{\rm\scriptsize ind\,}V}$ in \cite{CB}.
The semigroup consisting of these isometries $V$ will be denoted by
${\cal I(K)}$, and the distinguished automorphism corresponding to
$-{\bf1}\in{\cal I(K)}$ by $\gamma$. This automorphism turns ${\cal
  C(K)}$ into a ${\Bbb Z}_2$-graded algebra: ${\cal C(K)}={\cal
  C(K)}_0\oplus{\cal C(K)}_1$ and ${\cal C(K)}_g\cdot{\cal
  C(K)}_{g'}\subset{\cal C(K)}_{g+g'}$ where ${\cal C(K)}_g:=\{a\ |\ 
\gamma(a)=(-1)^ga\}$, $g\in{\Bbb Z}_2=\{0,1\}$.
\begin{prop}
  Let $V\in{\cal I(K)}$ with $\mbox{\rm ind\,}V=-1$, and let
  $k_V\in\ker V^*$ be unitary and skew-adjoint. Then
  $u_V:=\frac{1}{\sqrt{2}}({\bf1}+k_V)\in{\cal C(K)}$ is unitary, and
  \begin{equation}\label{1}
    \sigma_V(a):=u_V\varrho_V(a)u_V^*,\quad a\in{\cal C(K)}
  \end{equation}
  defines a unital *-isomorphism $\sigma_V$ from ${\cal C(K)}$ onto its
  even subalgebra ${\cal C(K)}_0$. 
\end{prop}
\begin{pf}
  (\ref{1}) clearly defines an injective *-endomorphism $\sigma_V$ of
  ${\cal C(K)}$. The direct sum decomposition ${\cal K}=\ker
  V^*\oplus\mbox{ran}\,V$ induces a ${\Bbb Z}_2$-graded tensor product
  decomposition ${\cal C(K)}={\cal C}(\ker V^*)\cdot{\cal C}
  (\mbox{ran}\,V)$, where the factors on the right are the subalgebras
  generated by $\ker V^*$ and $\mbox{ran}\,V$, respectively. From
  ${\cal C}(\ker V^*)_0=\Bbb C{\bf1}$ and ${\cal C}(\ker V^*)_1=\Bbb C
  k_V$ it follows that ${\cal C(K)}_0={\cal C}(\mbox{ran}\,V)_0+\Bbb
  C k_V\cdot{\cal C}(\mbox{ran}\,V)_1$. Since $u_Vau_V^*=a$ if
  $a\in{\cal C}(\mbox{ran}\,V)_0$ and $u_Vbu_V^*=u_V^2b=k_Vb$ if
  $b\in{\cal C}(\mbox{ran}\,V)_1$, $\mbox{ran}\,\sigma_V
  =u_V(\mbox{ran}\,\varrho_V)u_V^*=u_V{\cal C}(\mbox{ran}\,
  V)_0u_V^*+u_V{\cal C}(\mbox{ran}\,V)_1u_V^*={\cal C}(\mbox{ran}\,
  V)_0+\Bbb C k_V\cdot{\cal C}(\mbox{ran}\,V)_1={\cal C(K)}_0$.
\end{pf}
{\em Remarks.\/} 1. $V$ and $k_V$ as above always exist since $\dim{\cal
  K}=\infty$ and since $\ker V^*$ is invariant under complex
conjugation. $k_V$ is determined by $V$ up to a sign. The group of
Bogoliubov automorphisms $\{\varrho_U\ |\ U\in{\cal I(K)},\ 
\mbox{ind\,}U=0\}$ acts transitively on the set of
isomorphisms $\{\sigma_V\}$ by left multiplication.

2. The identification of ${\cal C(K)}$ with a `conventional' CAR
algebra $\cal A(H)$, generated by elements $a(f)$ and $a(f)^*$,
$f\in{\cal H}$ as in \cite{S}\footnote{$a(f)$ depends linearly on $f$
  in \cite{S}.}, depends on the choice of a closed subspace
(``polarization'') $\cal H\subset K$ such that $\cal K=H\oplus H^*$.
Given $\cal H$, one gets $\cal C(K)\cong A(H)$ by identifying
$f\in{\cal H}$ with $a(f)$.

3. In \cite{CB} we constructed left inverses $\Phi_V$ for Bogoliubov
endomorphisms $\varrho_V$. We get a left inverse $\varphi_V$ for
$\sigma_V$ by setting $\varphi_V(a):=\Phi_V(u_V^*au_V)$. Each
$a\in{\cal C(K)}$ has a unique decomposition $a=a_0+k_Va_1+b_1$ where
$a_g\in{\cal C}(\mbox{ran}\,V)_g$, $b_1\in{\cal C(K)}_1$, and then
$\varphi_V(a)=\varrho_V^{-1}(a_0+a_1)$. The conditional expectation
$\sigma_V\circ\varphi_V$ from ${\cal C(K)}$ onto ${\cal C(K)}_0$
coincides with the mean over $\Bbb Z_2$:
$\sigma_V(\varphi_V(a))=\frac{1}{2}(a+\gamma(a))$.

4. Since the representation theory of the CAR algebra is well
developed, one may use the isomorphisms from the Proposition to study
representations of the even CAR algebra. For example, in the algebraic
approach to the chiral Ising model \cite{I}, one has to study
representations of the form $\pi\circ\varrho_W$ with $\pi$ a Fock
representation of ${\cal C(K)}$ and $\varrho_W$ a Bogoliubov
endomorphism. It was observed \cite{Sz,JMB}, but regarded as a
curiosity, that the restriction of $\pi\circ\varrho_W$ to ${\cal
  C(K)}_0$ behaves like a representation $\pi\circ\varrho_{W'}$ of
${\cal C(K)}$ for some $W'\in{\cal I(K)}$ with $\mbox{ind\,}
W'=\mbox{ind\,}W-1$. This phenomenon is an immediate consequence of
the Proposition since $\pi\circ\varrho_W({\cal
  C(K)}_0)=\pi\circ\varrho_W\circ\sigma_V({\cal C(K)})\simeq
\pi\circ\varrho_{WV}({\cal C(K)})$ and $\mbox{ind\,}
WV=\mbox{ind\,}W-1$. In combination with results on representations of
${\cal C(K)}$ \cite{JMB,CB}, one obtains the decomposition of the
restriction of $\pi\circ\varrho_W$ to ${\cal C(K)}_0$ into irreducible
subrepresentations as discussed in \cite{JMB}.


\begin{thebibliography}{99}
\frenchspacing
\bibitem{S}
   St\o rmer, E.: The even CAR algebra, {\em Commun.\ Math.\ Phys.\ }{\bf16}
   (1970), 136--137.
\bibitem{A}
   Araki, H.: On quasifree states of CAR and Bogoliubov automorphisms,
   {\em Publ.\ RIMS Kyoto Univ.\ }{\bf6} (1970/71), 385--442.\\
   Araki, H.: Bogoliubov automorphisms and Fock representations
   of canonical anticommutation relations,
   in: {\em Operator Algebras and Mathematical Physics}, Am. Math. Soc.
   Vol. {\bf62} (1987), 23--141.
\bibitem{CB}
   Binnenhei, C.: Implementation of endomorphisms of the CAR algebra,
   {\em Rev.\ Math.\ Phys.\ }{\bf7} (1995), 833--869.
\bibitem{I}
   Mack, G. and Schomerus, V.: Conformal field algebras with quantum
   symmetry from the theory of superselection sectors, {\em Commun.\
     Math.\ Phys.\ }{\bf134} (1990), 139--196.\\
   B\"ockenhauer, J.: Localized endomorphisms of the chiral Ising
   model, {\em Commun.\ Math.\ Phys.\ }{\bf177} (1996), 265--304.
\bibitem{Sz}
  Szlach\'anyi, K.: Chiral decomposition as a source of quantum
  symmetry in the Ising model, {\em Rev.\ Math.\ Phys.\ }{\bf6}
  (1994), 649--671. 
\bibitem{JMB}
  B\"ockenhauer, J.: Decomposition of representations of CAR induced
  by Bogoliubov endomorphisms, DESY 94--173; An algebraic formulation
  of level one Wess--Zumino--Witten models, to appear in {\em Rev.\ 
    Math.\ Phys.}
\end{thebibliography}
\end{document}